\documentclass{emulateapj}
\usepackage{amsmath}
\usepackage{graphicx}
\usepackage{color}
\usepackage{amsmath}
\usepackage{graphicx}

\newcommand{\beq}{\begin{equation}}
\newcommand{\eeq}{\end{equation}}

\begin{document}

\title{Interferometric Measurement of Acceleration at Relativistic Speeds}
\author{Pierre Christian and Abraham Loeb\\
Astronomy Department, Harvard University, 60 Garden St., Cambridge, MA 02138}
\email{pchristian@cfa.harvard.edu; aloeb@cfa.harvard.edu}

\begin{abstract}
We show that an interferometer moving at a relativistic speed relative to a point source of light offers a sensitive probe of acceleration. Such an accelerometer contains no moving parts, and is thus more robust than conventional "mass-on-a-spring" accelerometers. In an interstellar mission to Alpha-Centauri, such an accelerometer could be used to measure the masses of exoplanets and their host stars as well as test theories of modified gravity.
\end{abstract}
\maketitle

\section{Introduction}
The Terrell effect  \citep{Terrell, Penrose} implies that a spherical source will always appear circular to a moving camera, in seeming contradiction to the naive expectation from Lorentz contraction in special relativity. In addition, for the simplest case of a sphere that extends across a small solid angle, the sphere will appear rotated.
 
Following this realization, much work has been done in generalizing the Terrell effect to cases where the solid angle is not necessarily small. It was found that while the spherical source is still viewed as a circle, more complex transformations than rotations are required to map the surface of the sphere to the one in the photographic plane \citep{Hollenbach, Suffern, Full3d}. Here we will restrict our attention to the simple case of a small solid angle, and focus on the temporal aspect of the Terrell effect. Even if the sphere is not featureless, one cannot infer whether the sphere is moving and Lorentz contracted or stationary but merely rotated just by taking a snapshot of it using a conventional camera. However, each point in the photographic plane can be traced back to the source through null geodesics. Due to the finite speed of light, geodesics corresponding to different points in the circular photograph travel for differing amounts of time. The timing information encoded by each light ray conveys the true nature of the sphere's motion. 

Whereas previous considerations of the temporal Terrell effect were based on an abstract construction involving a lattice of clocks \citep{Sheldon}, we propose a more natural way to detect the temporal Terrell effect based on interferometry, where the temporal information is encoded in phase measurements of light. In addition, we show that the temporal Terrell effect allows a measurement of acceleration with high precision due to a relative motion between the source and the camera.

Recently, an interstellar travel mission, \emph{Breakthrough Starshot} \citep{Starshot}, was initiated. The project aims to send a large number of gram mass microchips equipped with cameras to the nearest star system, Alpha-Centauri (4.2 light years away from Earth), at 20\% the speed of light. {\color{black} These microchips will aim to take close-up images of the planets in the star system. Furthermore, if during the travel these chips emit light pulses at known intervals, they can be used as standard-clock beacons much like pulsars. A passing gravitational wave will modulate the time between pulses, allowing the detection of long wavelength gravitational radiation.}

One other science benefit of such a mission could be to measure the masses of both exoplanets and their host stars as well as to test theories of modified gravity. Launching a spacecraft to a relativistic speed, however, requires a high acceleration that could potentially damage conventional mechanical accelerometers. A feasible alternative is to use the Doppler shift of the signal between the spacecraft and Earth to measure the accelerations. However, the total shift in speed for a spacecraft passing at 0.2c at a distance of 1 AU from an Earth mass planet is $\sim4 \times 10^{-3} \rm \; cm \; s^{-1}$. The best Doppler sensitivity for bright stars is 4 orders of magnitude weaker \citep{exoplanet} and it is difficult to imagine performance that is much better for a source that is much fainter than a star. 

In particular, an Earth-mass planet was recently discovered in the habitable zone (11.2 days orbit) around Proxima Centauri, a $0.12 M_\odot$ star in the three-star system of Alpha Centauri \citep{ExoCent}. The radial velocity technique by which the planet was detected allows to set only a minimum value on its mass, 1.3 Earth masses. Using a fleet of spacecrafts equipped with cameras and laser communication devices to the vicinity of this planet, \emph{Breakthrough Starshot} will aim to address the question whether the planet hosts life by taking color photographs and potentially measuring its mass.

Our proposed interferometric accelerometer possesses a sensitivity that is superior to the Doppler technique. Furthermore, due to the interferometric nature of the proposed accelerometer, an array of $N\gg 1$ cameras would provide $N^2$ independent measurements instead of $N$ if used merely for standard Doppler measurements.

The outline of the paper is as follows. In \S 2 we describe the Terell effect. In \S 3 we discuss the application of the temporal Terrell effect in the realm of interferometry. In \S 4 we propose a method of measuring accelerations using such an interferometer. {\color{black} In \S 5 we describe the uncertainties inherent in the measurements and in \S 6 we describe various scientific utilizations of such accelerometers.} Finally, we summarize our conclusions in \S 7.

\section{The Terrell Effect}
In the reference frame of a camera, located a distance $y_0 \gg R$ away from a radiating circle of rest frame radius $R$, the shape of the circular source is an ellipse given by 
\beq \label{eq:ellipse}
\gamma^2 (x - v t)^2 + y^2 = R^2 \; ,
\eeq
where $x$, $y$, $z$, and $t$ are the camera frame $4-$coordinates, $v$ the relative velocity between the camera and the source, and $\gamma$ is the corresponding Lorentz factor. Due to the finite speed of light, $c$, the emission time $t$ is given by 
\beq
t = T - \frac{y - y_0}{c} \; , 
\eeq
where $T$ is the arrival time of the photons to the camera. Without loss of generality, we can choose a time coordinate for which $T=0$, yielding,
\beq
\gamma^2 (x - v t)^2 + (y_0 - c t)^2 = R^2 \; .
\eeq
This equation can be solved for $t(x)$, the time at which the photons are emitted along the circle parameterized as a function of the $x$ coordinate,
\begin{align} \label{eq:master}
t(x) &= \frac{1}{c} \left\{ (y_0 + x \beta - y_0 \beta^2) \right. \nonumber
\\ &\left. \pm \sqrt{ (\beta^2-1) [ (x-y_0 \beta)^2 - R^2 ]  }   \right\} \; ,
\end{align}
where the $\pm$ in the square-root corresponds to light emitted from the far and near ends of the circle. The Terrell effect amounts to the fact that the $(x, y)$ values yielding a real $t$ solution inscribe a circle in the $(x,y)$ plane. However, the two end points of the circle corresponding to the equation
\beq \label{eq:endpoints}  %typo, the [
(\beta^2-1) (x-y_0 \beta)^2 - R^2  = 0 \; ,
\eeq
are rotated from that of a circle at rest by an angle (see Figure \ref{fig:circle}),
\beq
\theta_T = \arcsin \beta \; .
\eeq
%Rotation angle

\section{The Temporal Terrell Effect and Interferometry}
Next, we shift from the point of view of the camera to that of the source. We consider an interferometric array of cameras arranged in a circle of radius $R$ in its rest frame. This array is moving at a speed $c \beta$ in the $x$ direction at a large distance $y_0 \gg R$ from a point source. Because electromagnetism is symmetric with respect to time reversal, this set up is exactly the same as the Terrell configuration described in the previous section, where now the "camera" is the source and the "radiating circle" is the array.

\begin{figure}
  \centerline{\includegraphics[scale=0.26]{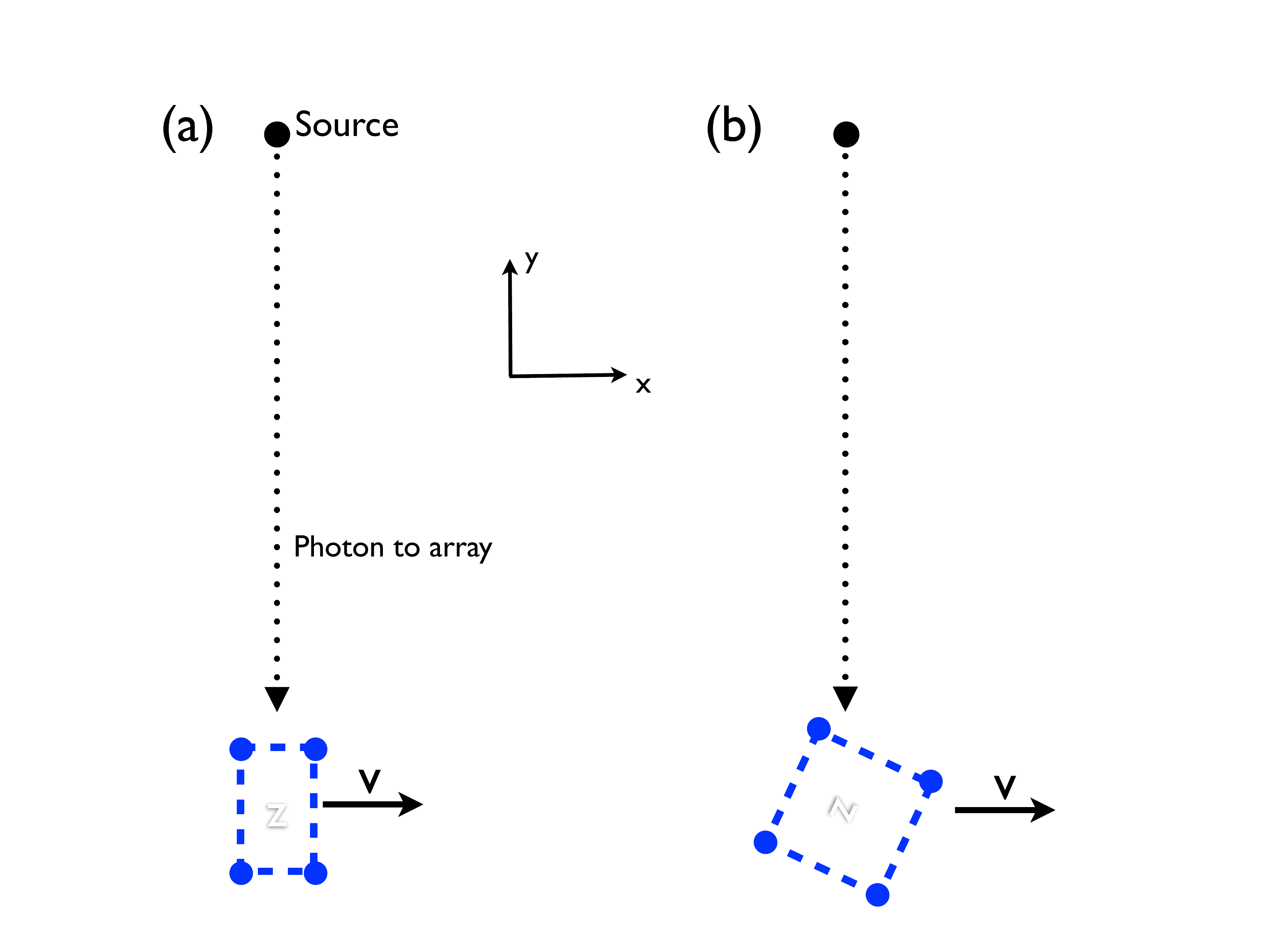}}
  \caption{\label{fig:square} The objective (a) and apparent (b) shape of a square interferometric array in the frame of the source. The square array is Lorentz contracted, resulting in the objective shape depicted in panel (a). However, due to the Terrell effect, the apparent shape of the array is rotated by an angle $\theta_T = \arcsin \beta$. If the array would take an image of the source, the moving source would appear as if it is located at a different angle on its sky due to relativistic aberation.} 
\end{figure}

In the frame of the point source, the array is distorted into the ellipse of equation (\ref{eq:ellipse}). However, due to the Terrell effect, the situation is equivalent to where the array stays circular but is rotated, as seen in Figure \ref{fig:square}. This is simply equivalent to relativistic aberration: a moving source will appear as if it is oriented at a different angle.

An advantage of an interferometer array over a "photon counting" camera is that it collects phase information. This allows us to leverage a lesser known aspect of the Terrell effect. A moving circle is mapped to a rotated circle; if the circle has no features this renders Lorentz contraction unphotographable. Temporal effects stemming from the Lorentz time dilation and the light travel time are, however, detectable by an interferometer. 

\begin{figure} 
  \centerline{\includegraphics[scale=0.3]{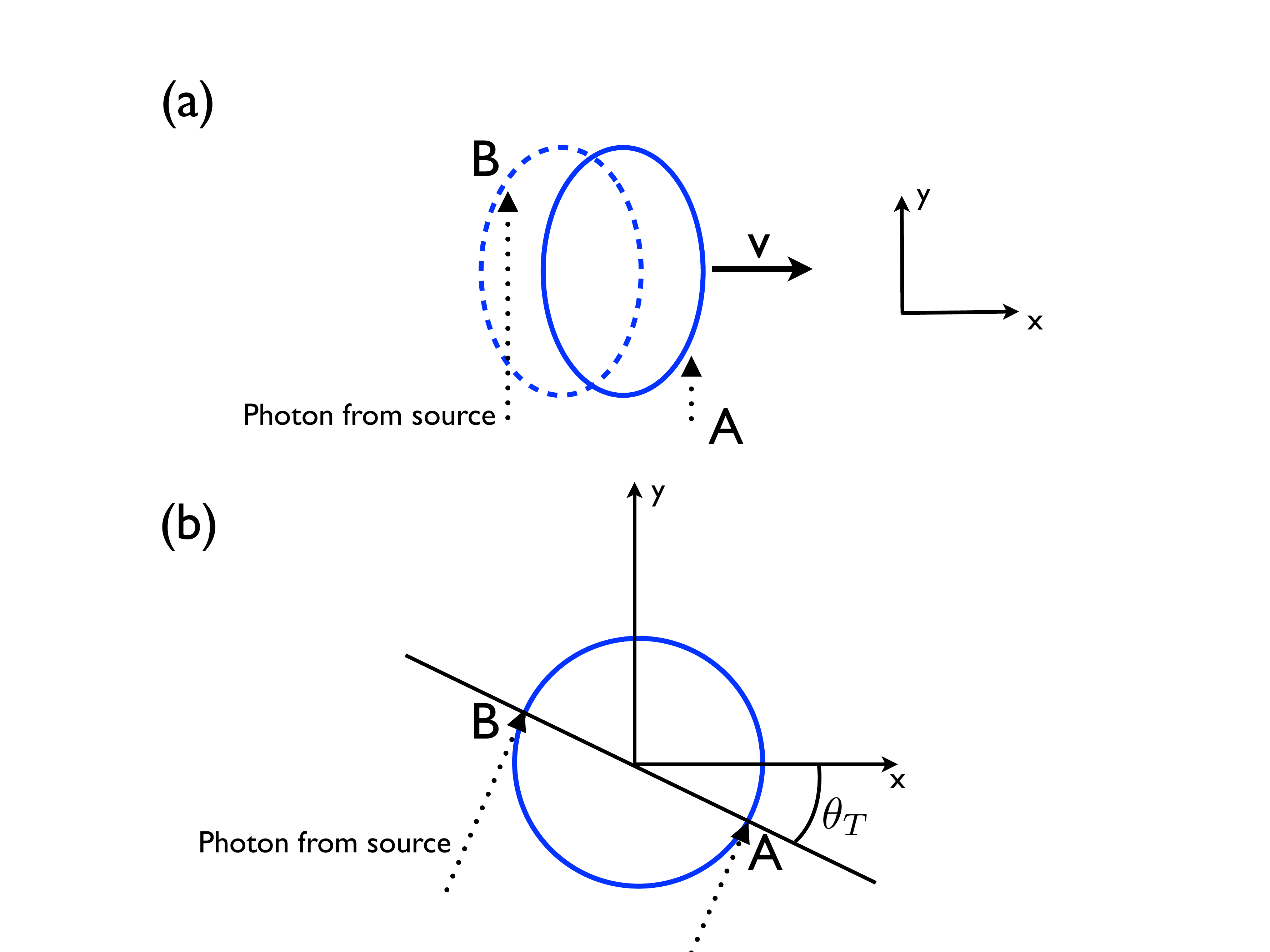}}
  \caption{\label{fig:circle} A circular array in the frame of the source (a) and its own rest frame (b). The dotted oval in panel (a) refers to the position of the array at an earlier time. In both frames, endpoints $A$ and $B$ satisfy equation (\ref{eq:endpoints}). In the source frame, these points label the first photon ($B$) and the last photon ($A$) received, forming the apparent shape of the array. Photons emitted after $B$ miss the array as it moves to the right, while photons emitted prior to $A$ miss the array as it is not there yet. In the frame of the array the photons are emitted at an angle $\theta_T = \arcsin \beta$. Every pair of points other than $A$ and $B$ reflect two solutions of equation (\ref{eq:master}), for positive and negative values of $y$ corresponding to the far and near sides of the array.} 
\end{figure}

The two end points of the circular array (where $t(x)$ has only one real solution; see Figure \ref{fig:circle}) are given by equation (\ref{eq:endpoints}), which can be solved to yield two equations for $x$ as a function of $y_0$ and $\beta$. Plugging these values of $x$ back to equation (\ref{eq:ellipse}) gives the arrival time of photons at these points if emitted from $y_0$ at $t=0$. Labelling the leftmost point A and the rightmost point B, we can calculate the time difference between the photons arriving at A and at B (see Figure \ref{fig:circle}),  
\beq
\Delta t = t_B - t_A =  \frac{2 \beta R}{c} \; .
\eeq
Photons arriving at B have to travel for an extra time $2 \beta R/c$ compared to photons arriving at A. The phase difference due to this extra travel time for electromagnetic waves with a frequency $\nu = \omega/2\pi$ and a wavelength $\lambda=c/\nu$ is
\beq \label{eq:phi}
\phi = \omega \Delta t = \frac{2 \omega \beta R}{c}  = 4 \pi \beta \frac{R}{\lambda} \; .
\eeq
For an interferometric array where $R/\lambda$ is of order unity, this extra phase is also of order unity at relativistic speeds with $\beta \sim 1$. However, the larger the array is (in units of wavelength), the larger would this extra phase be (modulus $2 \pi$, as with any phase measurements). Since the ratio $R/\lambda$ is known, this phase measurement can be used to infer the speed, $\beta$. Combined with a measurement of angular position change, one can use this to measure the distance between the array and the source, $y_0$. 

\section{The Terrell Accelerometer}
%Current accelerometers consist of mechanical systems that can be easily broken if subjected to large g-forces. If we are to send objects at relativistic speeds, it might be necessary for this object to suffer large amounts of acceleration, thus rendering on-board mechanical accelerometer problematic. We propose an antennae based accelerometer that can be used as long as the antennae remains usable. It contains no moving/mechanical parts, and is thus more robust than conventional accelerometers. 

Based on the derivations in \S 3, we are now at a position to consider an accelerometer on board a spacecraft that contains no mechanical parts and is usable as long as the spacecraft's antennae are operational. If our interferometric array passes close to an object, it would be subject to gravitational acceleration. This induces a time derivative, $\dot{\beta}$, to the array's speed that could be measured. Here, we will assume the mildly relativistic regime and expand all our equations to leading order in $\beta$. Moreover, we will assume that the acceleration is small compared to the velocity during the encounter, in the sense that $c\beta  \gg  |\dot{\beta} y_0/c|$. Furthermore, note that $\vec{\beta}$ is oriented initially radially away from the observer.

Since the Terrell effect respects the fact that information travels at most at the speed of light, $c$, in reality the gravitational field should also be expanded to leading order in $\beta$. In a relativistic theory, there is no action at a distance, and the gravitational changes are transmitted at speed $c$. Since the gravitational field scales as $1/d^2$, where $d$ denotes the distance between the source and the array, we find that to first order in $\beta$,
\beq \label{eq:grav_field}
\vec{g} = \left[ \frac{G M (\vec{n} - \vec{\beta}) }{ d ^2} \right]_{t_r} \; , 
\eeq
where $\vec{n}$ is the unit vector between the array and the object, $M$ the object's mass, $G$ is Newton's constant, and the equation is evaluated at the retarded time $t_r = (t - d/c)$. In electrodynamics, this is the familiar equation for the electric field of a moving body as given by the Lienard-Wiechert potentials, taken to first order in $\beta$ and setting the acceleration term to be small compared to the velocity term \citep{Jackson}. To lowest order, general relativity reduces to the gravitoelectric and gravitomagnetic fields. 

The $\vec{n}$ term in equation (\ref{eq:grav_field}) is the standard Newtonian acceleration, while the term proportional to $\vec{\beta}$ is a special relativistic term. In a flyby encounter, where the image is taken when the interferometric array and the gravitating object are at closest approach, $\vec{n} - \vec{\beta}$ evaluated at $t_r$ points directly towards the instantaneous position of the object and is perpendicular to the instantaneous $\vec{\beta}$.

For the sake of clarity, we examine the limit of Newtonian gravity, where $t_r \rightarrow t$ and only the $\vec{n}$ term is present. We consider a flyby where the array is passing the gravitating mass near closest approach, i.e. the $x$ component of the velocity is much larger than the $y$ component, $|\beta_x| \gg |\beta_y|$. Furthermore, we assume that the array is already near closest approach, where the $x$ displacement of the object to be much smaller than its $y$ displacement, $|x_0| \ll |y_0|$. In this case, the $x$ component of the acceleration is given as,
\beq \label{eq:betadot}
\dot{\beta}_x = \frac{G M}{c d^2} \frac{x_0}{y_0} \approx  \frac{G M}{y_0^3}  \beta t \; .
\eeq
The first derivative of equation (\ref{eq:phi}) yields, 
\beq \label{eq:phidot}
\dot{\phi} = 2 \pi \frac{R}{\lambda} \dot{\beta}_x \; ,
\eeq
yielding a relation between the rate of change of the phase and the acceleration of the system. This gives, 
\beq \label{eq:phidotgrav}
\dot{\phi} =  4 \pi \frac{R}{\lambda}   \frac{G M}{c d^2} \frac{x_0}{y_0}  \approx   4 \pi \frac{R}{\lambda}    \frac{G M}{y_0^3}  \beta_x t  \; ,
\eeq
As mentioned in section III, the value of $y_0$ can be inferred from combining the measurement of the phase, $\phi$, which directly probes $\beta_x \approx \beta$ with measurements of angular acceleration, $\dot{\theta}$, %alternative, $\beta$ is actually already known for large velocities, since it is just the velocity of the spacecraft if the velocity of the spacecraft is huge, since the body's inherent motion is nowhere close to 0.2c.
\beq
y_0 = \frac{c \beta }{\dot{\theta} } =\frac{c \lambda }{4 \pi R \dot{\theta}}\phi \; .
\eeq
The phase change due to the acceleration is given by the integral of equation (\ref{eq:phidotgrav}) with respect to time, 
\beq
\Delta \phi = 2 \pi  \frac{R}{\lambda}    \frac{G M}{y_0^3} \beta_x (\tau-\tau_0)^2 \; ,
\eeq
where $\tau$ is the observation time and $\tau_0$ is an arbitrary starting time. This phase change is parabolic with respect to time; it is negative as the spacecraft approaches the gravitating body, and positive as it moves away from it. If we define $T_0=0$ at the point where the phase change is zero (when the spacecraft is directly in front of the gravitating body), then the mass of the body is given by
\beq
M = \frac{\lambda}{R} \frac{y_0^3}{G M} \frac{2 \pi}{ \beta_x T^2} \Delta \phi \; .
\eeq
%Thus, given the $x$ displacement, one can measure the mass of the gravitating object, $M$. 
%Practically, the time derivative $\dot{\phi}$ gives a direct measurement of the mass,
%\beq
%M = \frac{\ddot{\phi}}{G \beta} \left( \frac{c \lambda}{4 \pi R} \right)^2 \frac{\phi^3}{\dot{\theta}^3} \; .
%\eeq
For an array of $N$ antennae, this procedure can be repeated for every single baselines in the array, resulting in $N(N-1)/2$ independent measurements of the mass, $M$. 

\section{Noise and systematics}
To assess the signal's detectability, we compare them to the noise inherent in the interferometer. {\color{black}Free from atmosphere induced errors plaguing ground based interferometers, the fundamental} phase noise of the interferometer is \citep{TMS},
\beq \label{eq:noise}
\sigma_\phi = \frac{T_S}{T_A \sqrt{2 B \tau N}}\; , 
\eeq
where $T_S$ and $T_A$ are respectively the system and antenna temperatures, $B$ the bandwidth, $N$ the number of baselines, and $\tau$ the observation time. The system temperature is given in terms of the physical temperature $T$ of the detector by \citep{TMS},
\beq
T_S = T \left[ \frac{   \frac{h \nu}{ k_B T}   }{ e^{\frac{h \nu}{ k_B T} } -1 } \right] \; ,
\eeq
where $h$ and $k_B$ are the Planck and Boltzmann constants, respectively. The antenna temperature $T_A$ is related to the source's intensity by
\beq
T_A = \frac{A S}{2 k_B} \;,
\eeq
where $A$ is the area of the antenna and $S$ the flux density of the source.  For our purposes, the point source chosen for the measurement is the Sun (as seen from the Alpha-Centauri system), $N$ is taken to be $100$, $A$ to be $1$ cm$^2$, the physical temperature $T$ to be the temperature $1$ AU from a sunlike star ($270$ K), and the bandwidth $B$ to be 10 percent of the frequency of observation. {\color{black} For such parameters, equation (\ref{eq:noise}) implies a phase noise level at a wavelength of $1 \mu$m of 
\beq \label{eq:noise}
\sigma_\phi \approx 3\times 10^{-22}\sqrt{\frac{1 \; \rm{s} }{\tau N}}   \;.
\eeq}
{\color{black}Substituting the passage time $\tau \approx d/\beta c$ as the observation time,  we obtain that $\sigma_\phi \approx 10^{-22}$ for $d$ being the Earth-Moon distance.} 

{\color{black} Within a single chip, thermal expansion causes the length of the baselines to change. Baselines between antennae located onboard different spacecrafts could also change due to orbital drifts. These phenomena generates systematic uncertainties that could be controlled by monitoring the relative positions of the antennae. This can be done by transmitting light signals between the antennae. Timing these signals allows for the distances and relative velocities between antennae to be known. In the swarm configuration where each antennae is on board a different spacecraft, either laser or the same transmission that is used to communicate back to Earth can be utilized for this purpose. In the case of multiple antennae on a single chip, optical fibers can be used as an alternative.}

\section{Science utilization}
\subsection{Weighting Stellar and planet masses}

At a distance of $1$ AU from a Sun-like star, the total change of phase for an array moving at $\beta = 0.2$ is given by
\beq \label{eq:stareq}
\Delta \phi \approx 3\times10^{-7} \times \left(\frac{R}{\lambda}\right) \times \left( \frac{1 \rm{AU}}{d} \right) \times \left( \frac{M}{M_{\odot}} \right) \; ,
\eeq
where in this case we equated the observation time to the crossing time, $\tau \approx d/\beta c$. For an array of antennae onboard the spacecrafts envisioned for the Breakthrough Starshot initiative with $R \sim 1$ m and $\lambda \sim 1 \mu$m, this phase change is $\Delta \phi \sim 0.3$, {\color{black} which is large compared to the noise described by equation (\ref{eq:noise})}. One could also send many antennae on separate spacecrafts, and the interferometry process can be conducted across different spacecrafts. In this case, $R$ would be the distance between spacecrafts, allowing the use of longer wavelengths to produce the same amount of phase change. %For example, for an interferometric array with a baseline of $R\sim10^2$ cm, one can obtain a phase change of order unity with $\lambda \sim 1$ $\mu$m.

If the spacecraft passes sufficiently close to a planet, the gravitational pull of the planet would dominate over that of its parent star. For a planetary system similar to the Sun-Earth system, this would occur when the planet-spacecraft separation is about $3 \times 10^{10}$ cm, or roughly the Earth-Moon distance. In such a flyby, the total change of phase at $\beta = 0.2$ is given by 
\beq \label{eq:planeteq}
\Delta \phi \approx 4.7\times10^{-10} \times \left(\frac{R}{\lambda}\right) \times \left( \frac{3\times10^{10} \rm{cm}}{d} \right) \times \left( \frac{M}{M_{\oplus}} \right) \; ,
\eeq
which gives $\Delta \phi \sim 4.7 \times 10^{-4}$ for the configuration with $R \sim 1$ m and $\lambda \sim 1 \mu$m. {\color{black} This phase difference is large compared to the noise described by equation (\ref{eq:noise}). This will complement an alternative measurement of the planet's mass that is enabled by sending a spacecraft to the system, such as by determining the orbital inclination of the planet by resolving the planet's orbits as the spacecraft approaches the planetary system.}

We note that it is not necessary to use the star or planet as the light source. If the array is focused on some distant point source like the Sun, the acceleration of the detector will still be apparent. However, if the star or planet is used as the light source, its resolved image can be viewed as a collection of point sources, and each of them will suffer the same temporal drift in phase due to the acceleration. The observer would see each pixel in the u-v plane of the interferometer drifting in time. This will also enable a measurement of the acceleration.

{\color{black}
\subsection{Measuring density profiles of planetary systems}

Approaching the target star system with antennae allows one to measure the enclosed mass of the system as a function of distance from the star. In particular, this allows to measure the cumulative mass of Kuiper belt analogs in the target system. At a distance of $30$ AU,  a Kuiper belt analog of combined mass $\sim 30 M_\earth$ produces an extra acceleration of 
\beq
c \dot{\beta} = \frac{30 G M_\earth}{(30 \;\rm{AU})^2} \; ,
\eeq
which translates into an additional phase change of
\beq
\Delta \phi \approx 10^{-12} \times \left(\frac{R}{\lambda}\right) \times \left( \frac{30 \rm{AU}}{d} \right) \times \left( \frac{M}{30 M_{\earth}} \right) \; . 
\eeq

For the configuration with $R \sim 1$ m and $\lambda \sim 1\mu$m, this phase change becomes $\sim 10^{-6}$, which is above the noise implied by equation (\ref{eq:noise}).
 }
 
\subsection{Measuring the Milky Way mass}

Travel to the closest star from Earth requires an extended period of time for the journey. At $\beta=0.2$, it will take roughly $20$ years to travel from Earth to Alpha-Centauri. During this travel time, one can use the phase change of any source to weight the Galactic mass. The current estimate for the enclosed Milky Way mass at the distance of $8$ kpc from the Galactic center is $\sim 10^{11} M_\odot$ \citep{Reid}; therefore,
\beq
\Delta \phi \approx 10^{-30}  \times \left( \frac{R}{\lambda} \right) \times \left( \frac{T}{\rm sec} \right)^2 \; .
\eeq
Although this number is small, the travel time is longer than in the case considered in \S 4.1. If measurements are taken for the full travel time of $\sim20$ years to Alpha-Centauri, the total phase change is,
\beq
\Delta \phi \approx 3 \times 10^{-15} \left(\frac{R}{\lambda}\right) \; .
\eeq
This will provide a phase change of unity for $\lambda \sim 1 \mu$m if $R \sim 10^{9}$ cm. While the phase change for the $R \sim 1$ m configuration is small, note that it is still large compared to the noise described by equation (\ref{eq:noise}) provided that the Sun could be used as the point source in the experiment. 

{\color{black}
\subsection{Testing modified gravity}
Long distance modifications of gravity could be constrained by monitoring the acceleration of the spacecraft \citep{modgrav1, modgrav2, modgrav3}. These modifications could be parameterized by adding a Yukawa term to the usual $1/r$ gravitational potential,
\begin{align}
\Phi(r) &= -\frac{G M}{r} \left[ 1 + \alpha e^{-r/ l} \right] 
\\ &\approx - \frac{G M}{r}  \left[1 + \alpha - \frac{\alpha r}{l} + \frac{r^2}{2 l^2} + \ldots \right] \; ,
\end{align}
where $\alpha$ is the strength of the Yukawa force and $l$ is the characteristic length-scale at which the gravitational force is substantially modified from $1/r^2$. As noted in \cite{modgrav1}, the zeroth order correction is simply a rescaling of the mass and the first order correction is a constant shift in the potential. The lowest order observable term is second order in $(r/l)$ and produces the constant acceleration
\beq
\delta a = \frac{\alpha G M}{2 l^2} \; ,
\eeq
where we use $\delta$ to signify the difference in physical quantities between the cases with and without the Yukawa interaction. Modifying equation (\ref{eq:betadot}) by this term yields the difference between the $x$-accelerations of $1/r^2$ and Yukawa gravity to be 
\beq
\delta \dot{\beta}_x = \frac{\alpha G M }{c 2 l^2} \frac{x_0}{y_0} \approx \frac{G M }{l^2 y_0} \beta t \; .
\eeq
This produces a total phase shift of, 
\beq
\delta \phi = 2 \pi \frac{R}{\lambda} \left(\frac{\alpha G M}{l^2} \right) \frac{d}{\beta c^2} \; , 
\eeq
where $d$ is the distance to the source and again we equated the observation time to the crossing time, $d/\beta c$. Placing the detectors at multiple distances from a star, the strength of the Yukawa potential, $\alpha$ can then be directly measured. 
The smallest phase shift measureable is given by (\ref{eq:noise}), implying that the experiment is sensitive to measure $\alpha$ down to,
\beq
\alpha_{min} = \sigma_\phi \frac{l^2}{G M} \left( \frac{\beta c^2}{d} \right) \frac{\lambda}{2 \pi R} \; .
\eeq
For a Yukawa potential with length-scale $l \sim 100$ AU, a detector located at $y_0 \approx l$ around a sun-like star can measure $\alpha$ down to 
\beq
\alpha_{min} \approx 2 \times 10^{-17} \times \sqrt{\frac{100}{N}} \times \frac{\lambda}{R} \; ,
\eeq
where $N$ is the number of baselines in the measurements. With $10$ antennae, the number of baselines becomes $N\approx100$, allowing the measurement of $\alpha_{min}$ to a few orders of magnitudes better than state of the art experiments \citep{modgrav1}. In addition to its sensitivity, another benefit for using this method is that one can test the long range modification of gravity on much larger scales than previously possible.  
}

\section{Conclusion}
We showed that the temporal Terrell effect provides an exquisite probe of acceleration for an interferometric array traveling at a relativistic speed. In the context of the envisioned parameters of the Breakthrough Starshot spacecraft ($R\sim 0.1-4$ m, $\lambda \sim 1 \mu$m, $\beta \sim 0.2$), equations (\ref{eq:stareq}) and (\ref{eq:planeteq}) imply measureable phase variations for a Sunlike star and an Earth mass planet, respectively. An interferometric array of $N$ elements onboard such a spacecraft could provide $N (N-1)/2$ measurements of the mass, allowing a new observational method for measuring masses  and testing theories of modified gravity in astronomy.  

\section{Acknowledgement}
The authors thank Michael Johnson and Martin Rees for comments on the manuscript {\color{black}and an anonymous referee for insightful suggestions}. This work was supported in part by a Starshot grant from the Breakthrough Prize Foundation to Harvard University and by NSF grant AST-1312034.

\end{document}